# Event-driven Spectrotemporal Feature Extraction and Classification using a Silicon Cochlea Model

Ying Xu, Samalika Perera, Yeshwanth Bethi, Saeed Afshar, André van Schaik
International Centre for Neuromorphic Systems, The MARCS Institute, Western Sydney University, Kingswood, NSW 2751, Australia
ying.xu@westernsydney.edu.au

*Abstract*—This paper presents a reconfigurable digital implementation of an event-based binaural cochlear system on a Field Programmable Gate Array (FPGA). It consists of a pair of the Cascade of Asymmetric Resonators with Fast Acting Compression (CAR-FAC) cochlea models and leaky integrate-and-fire (LIF) neurons. Additionally, we propose an event-driven SpectroTemporal Receptive Field (STRF) Feature Extraction using Adaptive Selection Thresholds (FEAST). It is tested on the TIDIGTIS benchmark and compared with current event-based auditory signal processing approaches and neural networks.

1 INTRODUCTION

In the human auditory pathway, information is extracted and conveyed through sequences of action potentials, or spikes. The spike streams form robust representations that are important for perception. The human sensory system achieves real-time, low-power, and noise-robust performance while operating in such an asynchronous "event"-based way. To mimic the efficiency of signal processing in the human auditory system, biologically inspired auditory sensors and algorithms have been implemented and investigated. For example, (Liu, van Schaik, Minch, & Delbruck, 2010, 2014) developed a 2×64×4 channel dynamic audio sensor that used an analogue cascade filter bank and pulse-frequency modulated circuits to emulate the peripheral auditory system and auditory nerve to generate spike streams; (Yang, Chien, Delbruck, & Liu, 2016) used a synchronised delta modulator to generate audio events; (Singh et al., 2018) developed a digital multi-rate cochlea model on FPGA where a digital leaky integrate-and-fire (LIF) neuron model with different thresholds was used to model auditory neurons of the human auditory system with different thresholds.

Such neuromorphic auditory sensors encode acoustic information into spikes in real-time at a low data rate, which make them an ideal solution for real-world applications. In recent decades, efforts have been made to investigate neuromorphic sensing approaches to extract acoustic features from auditory spikes. For example, it has been argued that statistical features embedded in spike streams could be the mechanism for the precise encoding of auditory cues that are important for recognition (Gerstner & Kistler, 2002). Therefore rate-code based features (Neil & Liu, 2016), inter-spike interval distributions (Li, Delbruck, & Liu, 2012), (Uysal, Sathyendra, & Harris, 2006) inter-spike velocity (Chakrabartty & Liu, 2010), and exponential features (Anumula, Neil, Delbruck, & Liu, 2018) have all been investigated in speaker identification and speech recognition tasks. (Rasetto, Dominguez-Morales, Jimenez-Fernandez, & Benosman, 2021) proposed a feature extraction approach to extract spectrotemporal features from a cochlea model built with "event"-based filters for a command recognition task.

In addition to neuromorphic auditory data processing, event-driven feature extraction algorithms have been more widely investigated in neuromorphic vision systems. With the increase in the adoption of neuromorphic vision sensors, various dense tensor representations for the sparse asynchronous event data have been proposed and investigated to learn the spatiotemporal features (Afshar, Nicholson, van Schaik, & Cohen, 2020; Baldwin, Liu, Almatrafi, Asari, & Hirakawa, 2022; Maqueda, Loquercio, Gallego, Garcia, & Scaramuzza, 2018).

In (Afshar, Nicholson, et al., 2020; Cohen et al., 2019), the event-based time surface representations for event-based vision data have been used in extracting features for a range of tasks, such as object recognition on unmanned aerial vehicles (UAVs) (Zappa et al., 2020) and single photon avalanche diode (SPAD) sensors data processing (Afshar, Hamilton, Davis, van Schaik, & Delic, 2020).

In (Afshar, Ralph, et al., 2020), Feature Extraction using Adaptive Selection Thresholds (FEAST) was proposed for event-based vision data using the time surfaces representation. The FEAST method has been investigated for a range of applications such as object tracking (Ralph et al., 2022), event-based supervised learning (Bethi, Xu, Cohen, van Schaik, & Afshar, 2022) and inspired activity-driven adaptation in spiking neural networks (SNNs) (Haessig et al., 2020).

*Correspondence: Ying Xu, International Centre for Neuromorphic Systems, The MARCS Institute, Western Sydney University, Locked Bag 1797, Kingswood, NSW 2751, Australia. E-mail: ying.xu@westernsydney.edu.au

To investigate spectrotemporal representations for event-based auditory data, in (Xu, 2019), the FEAST method was investigated in audio to extract spectrotemporal features for an isolated spoken digits recognition task and showed improved performance. In this work, we extend the work and propose to use FEAST to build a computational auditory cortical model – the Spectrotemporal Receptive Field (STRF) model. The proposed event-driven STRF approach is applied to the binaural cochlear system for a multi-resolution spectrotemporal analysis.

## 2 MATERIALS AND METHODS

### 2.1 THE EVENT-BASED BINAURAL CAR-FAC SYSTEM ON FPGA

In the previous work, we implemented a digital cochlea model, the Cascade of Asymmetric Resonators with Fast Acting Compression (CAR-FAC) cochlea model (Lyon, 2017) on a Field Programmable Gate Array (FPGA) for sound localisation (Xu et al., 2021). This model approximates the physiological elements that make up the human cochlea, including the basilar membrane (BM), the outer hair cells (OHCs) and the inner hair cells (IHCs), as shown in Figure 1, and mimics its qualitative behaviour. The digital cochlea is reconfigurable in filter parameters and channel numbers. This work extends the cochlea model to an event-based binaural cochlear system. It includes a CAR-FAC cochlea pair and LIF neurons to generate auditory spike streams.

The architecture of the event-based binaural cochlear system is shown in Figure 2. Each "ear" in the system implements the components of the CAR, the digital OHC (DOHC), the digital IHC (DIHC), the automatic gain control (AGC), the lateral inhibition (LI), and the LIF neuron. One "ear" can be switched off so that the system operates as a single CAR-FAC model. The FAC part that introduces nonlinearities can also be switched off so that the system operates as a linear CAR model. The details of the CAR-FAC module were described in (Xu, Thakur, Singh, Wang, & van Schaik, 2016), (Xu, Thakur, et al., 2018), and (Xu et al., 2019). The LIF neuron here is implemented using:

$$LIF[s,t] = LIF[s,t-1] + c_{LIF} \times (IHC_{LI[s,t]} - LIF[s,t-1]) \quad (1)$$

$$c_{LIF} = 1/(f_s \times \tau_{LIF}) \quad (2)$$

where $LIF[s,t]$ is the membrane potential of a neuron connecting to the CAR-FAC LI output, $IHC_{LI[s,t]}$ in channel $s$ at time $t$, $f_s$ is the sampling frequency, and $\tau_{LIF}$ is the time constant of the LIF neuron. When $LIF[s,t]$ is above a *threshold*, a spike, $spk[s,t]$, is generated at time $t$ in channel $s$:

if $LIF[s,t] > threshold$:

$$spk[s,t] = 1,$$

$$LIF[s,t] = V_{reset}$$

else:

$$spk[s,t] = 0 \quad (3)$$

and the membrane potential of the neuron is reset to value $V_{reset}$. The generated spike streams encode the amplitude of each channel response that is used in the following feature extraction.

### 2.2 UNSUPERVISED FEATURE EXTRACTION

#### 2.2.1 Unsupervised Feature Extraction Using Adaptive Selection Thresholds (FEAST)

The FEAST method in (Afshar, Ralph, et al., 2020) extracts spatio-temporal features for event-based vision data using real-valued exponentially decaying kernels and 2-D "neurons". The use of exponentially decaying kernels for event-based processing was described in (Tapson, Cohen, & van Schaik, 2015) and called a "time surface" in (Lagorce, Ieng, Clady, Pfeiffer, & Benosman, 2015). The time surface is generated by applying an exponential decay with a time constant $\tau_v$ on a local (typically square) neighbourhood centred on the current event. For example, an event $e_i$ occurred at time $t$:

$$e_i = [x, y, t, p]^T \quad (4)$$

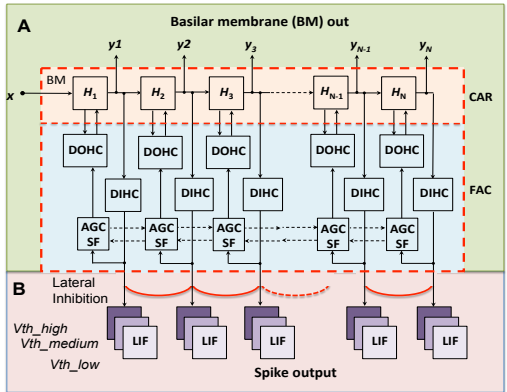

Figure 1 Structure of the CAR-FAC model (A) and the LIF neurons preceded by lateral inhibition (B). $x$ is the input sound, $H_1$ to $H_N$ are the transfer functions of the CAR part, and $y_1$ to $y_N$ represent the CAR-FAC Basilar Membrane (BM) output. The characteristic frequencies (CFs) of the CAR resonators decrease from left to right. The DOHC, the DIHC and the AGC loop comprise the FAC part. Each DIHC is connected to nine LIF neurons with three thresholds, *Vth_low*, *Vth_medium* and *Vth_high*, after a Lateral Inhibition function between neighbouring channels.

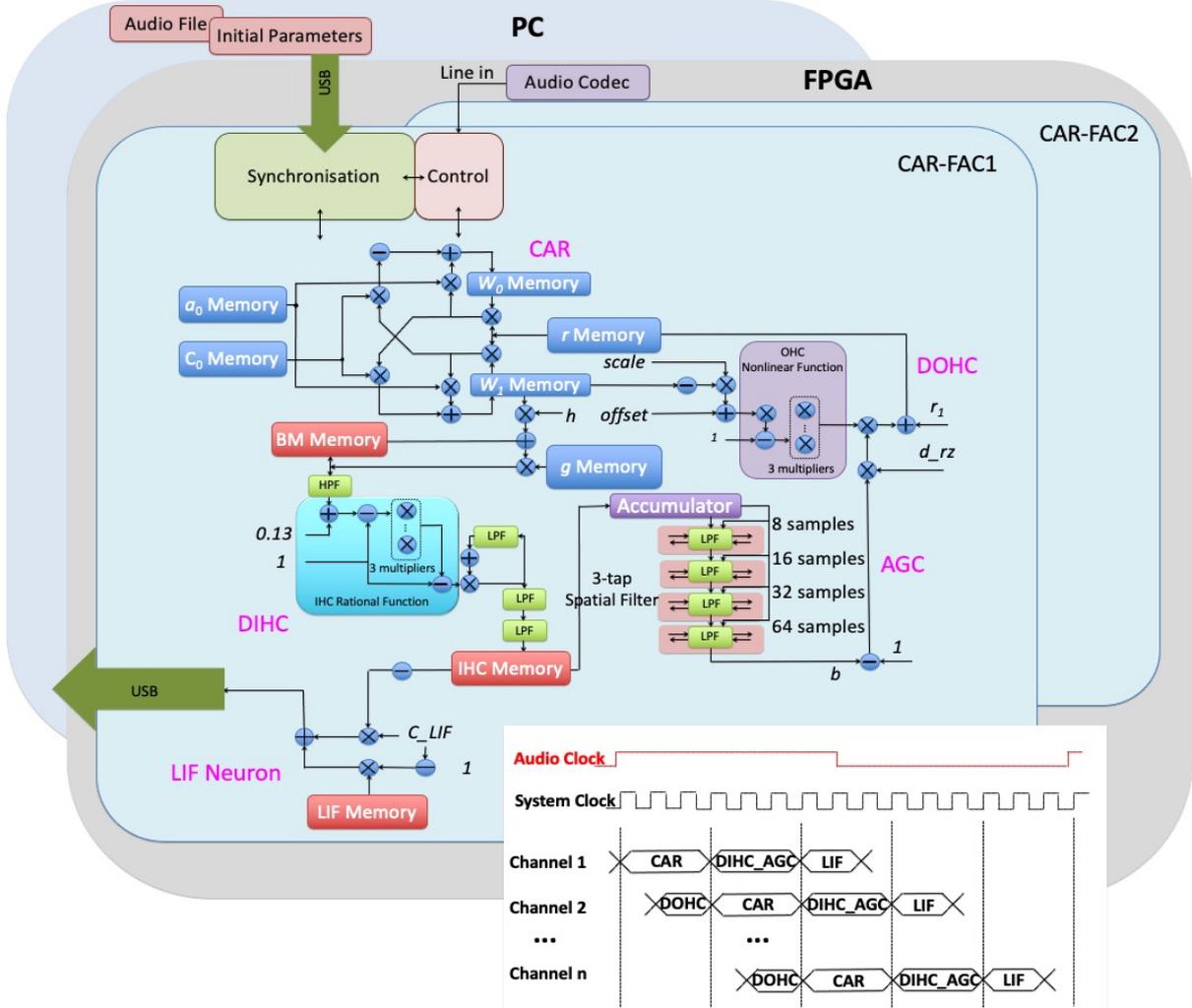

Figure 2 Architecture of the binaural CAR-FAC FPGA system. The system consists of an audio codec and two "ears". Each of the ears includes a CAR-FAC module, a controller module, and an interface module. The FPGA board is hosted by a PC through a USB interface. The inset shows the system timing diagram demonstrating the pipelined CAR-FAC. With time multiplexing and pipeline techniques, a binaural real-time n-channel CAR-FAC system is built using only one CAR-FAC module and one LIF module for each ear.

where $x, y$ represent the spatial location of the pixel with reference to the event-based sensor and $p \in \{-1,1\}$ is the polarity of the event.

The time surface $S_i(\mathbf{u}, p)$ at the location ($\mathbf{u} = [x_i, y_i]^T$) of the event $e_i$ at time $t$ can be calculated as:

$$S_i(\mathbf{u}, p) = e^{-(t_i - \Gamma_i(\mathbf{u},p))/\tau_v} \quad (5)$$

where $\Gamma_i(\mathbf{u}, p)$ is the timestamp of the latest event that occurred at the location $u$. The time surface of a pixel in the spatial neighbourhood of size $R$ around an event location is considered as an event context (E_C) (of size $(2 \times R + 1) \times (2 \times R + 1)$). In event-based vision, the local E_C describes the recent time history of events in the spatial neighbourhood of an event in 2-D.

FEAST learns spatiotemporal features from the E_Cs through 2-D "neurons". Each neuron has randomly generated initial threshold and weights. The neurons act as feature extractors with individual adaptive thresholds via a competitive strategy. A similarity measure, $\cos(\theta)$, between the E_C and the neuron's weights is used as a metric to match the event contexts with the weights of each neuron:

$$similarity = \cos(\theta) = \frac{E\_C \cdot W_i}{||E\_C|| \cdot ||W_i||} \quad (6)$$

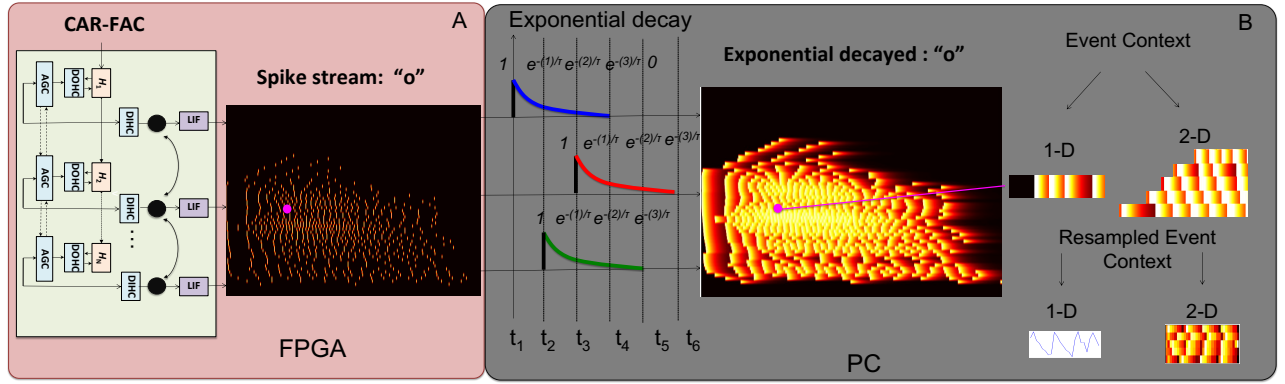

Figure 3 Construction of the auditory time surface window. (A) Spike streams generated from the binaural CAR-FAC system on FPGA; (B) When a spike occurs in a frequency channel, the value decays exponentially as t increases, forming the exponential decaying feature. A 1-D/2-D event context (E_C) is generated on a local neighbourhood centred on the current event (in magenta). The 1-D E_C provides temporal feature extraction, and the 2-D E_C provides spectrotemporal feature extraction. The E_Cs are resampled to form a uniform size for all the events as described in the text.

where $W_i$ denotes the weights of neuron $i$, and $\frac{E\_C}{||E\_C||}$ and $\frac{W_i}{||W_i||}$ are the normalisations of E_C and the weights. After normalisation, the similarity is calculated as a dot product of normalised E_C and weights.

In the learning phase, each neuron's unique threshold acts as a selection boundary. The neuron with the highest similarity that also crosses its selection threshold is picked as the winner neuron, which then emits a spike. The thresholds of the neurons are dynamic during learning, and are adapted based on two rules:

1. If there is a winner, then increase the threshold $Vth_i$ of neuron $i$ by a fixed amount $\Delta I$.
2. If there is no winner, then decrease all the neurons' thresholds by a fixed amount of $\Delta E$.

The E_C is then used to update the winning neuron's weights with a fixed mixing rate as follows:

$$W_{i\_update} = (1 - \eta) \times W_i + \eta \times E\_C \qquad (7)$$

Where the weights $W_i$ of neuron $i$ to which the E_C is successfully matched, and $\eta$ is the mixing rate used to update the weights of the neuron. The weights of the neurons form features that cover the feature space of the input signals. The use of a dynamic threshold ensures that the rate of firing of all neurons is approximately equal across the dataset, as increasing the threshold on the matching feature serves to specialise each neuron from other neurons. If the weights are coding poorly for the incoming feature, then the global threshold decrease serves to expand the range of input features to which the neurons will respond. This learning process is dynamic and responsive to the statistics of the incoming data.

When the FEAST is applied to the event-based audio data, the E_C needs to be formed differently. Figure 3 shows the construction of the E_Cs and the details will be illustrated in the next two sections. In this paper, we use the FEAST to build the event-based multi-resolution spectrotemporal analysis. The computational SpectroTemporal Receptive Field (STRF) model is inspired by psychoacoustical and neurophysiological findings in the early and central stages of the auditory system (Chi, Ru, & Shamma, 2005). The model provided a unified multi-resolution representation of the spectral and temporal features likely critical in the perception of sound. It mimics aspects of the responses of higher central auditory stages, especially the primary auditory cortex. Functionally, it estimates the spectral and temporal modulation content of the auditory spectrogram via a bank of filters that are selective to different spectrotemporal modulation parameters ranging from slow to fast rates temporally, and from narrow to broad scales spectrally (Chi et al., 2005). Here we break the proposed event-based spectrotemporal feature extraction into two steps:

### 2.2.2 Temporal Feature Extraction –1-D FEAST

The CAR-FAC model shows highly frequency-dependent gains and the connecting LIF neurons encode the amplitude of the channel responses. A similar amplitude coding is also used by (Liu et al., 2014). Figure 4 show the CAR-FAC response to a TIDIGITS utterance "o". In the middle frequency channel, 650 Hz, the response shows the highest gain in amplitude, and thus higher spike numbers than the higher (1000 Hz) and lower frequency channels (180 Hz). Additionally, the inter-spike interval encodes the changes in amplitude. For example, for an increment in amplitude, the spike train shows a gradually decreasing inter-spike interval, whereas, for a decrease in amplitude, the spike train shows a gradually increasing inter-spike interval. In this way, the spike trains of each channel encode syllabic rates of speech. In speech and music, there are three kinds of temporal modulations (Chi et al., 2005) in the cochlear outputs. Slow modulations that reflect the syllabic rates of speech. They are superimposed upon the intermediate rate modulations due to inter-harmonic interactions occurring at a rate that reflects the fundamental frequency of the input, which in turn are riding upon the fast frequency component driving this channel best, the characteristic frequencies (CF) of each cochlear channel.

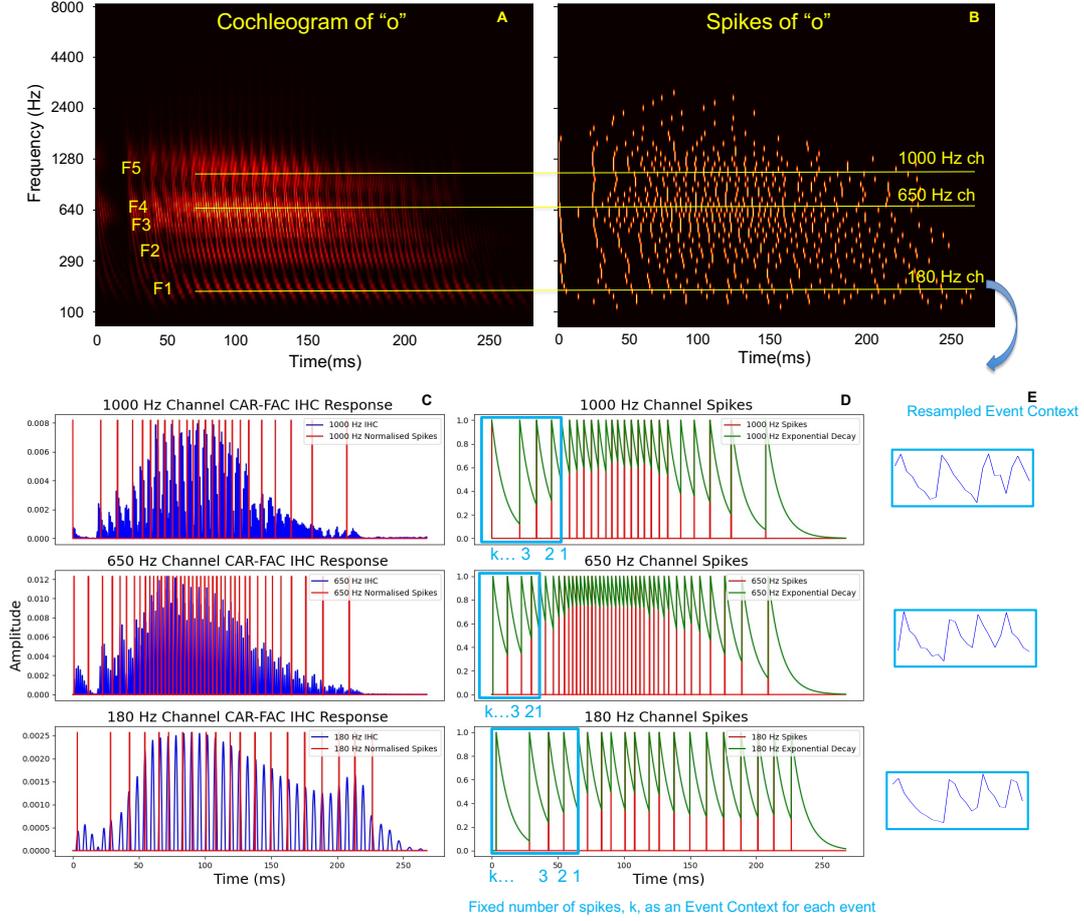

Figure 4 The CAR-FAC output of utterance "o", (A) formants (F1-F5) are labelled and three channels, 180 Hz, 650 Hz, and 1000 Hz, are highlighted in yellow; (B) the spikes of utterance "o"; (C) the IHC output from CAR-FAC and the generated spikes encoding amplitude of the three highlighted channels; (D) the time surfaces in these channels; (E) the resampled E_Cs that preserves temporal information.

The first step of Event-based Spectrotemporal Feature Extraction, 1-D FEAST, is to extract such syllabic rates, or slow amplitude changes, from each cochlear channel temporally:

As shown in Figure 4 (C) and (D), we apply an exponential kernel decaying with a time constant $\tau$ on each event across the channels:

$$\tau = f_s \times 10^{-3} \qquad (8)$$

where $f_s$ is the sampling frequency, and $\tau$ determines a duration over which the previous event has an impact on the scene, and the current event represents the highest energy, 1.0, as shown in Figure 4 (D). We then define a 1-D E_C for each event that includes a fixed number of spikes. Each E_C should include a sufficient number of spikes such that a change in amplitude can be represented. Since the E_C generated for each spike has a different duration in time, or number of samples, we then resample the E_C into a fixed number of samples. After resampling, all the E_Cs have a same number of samples, while preserving the encoded temporal features. For example, in Figure 4 (E), an onset is shown in five consecutive spikes with gradually increased inter-spike intervals in all the channels.

FEAST is then applied to the 1-D E_Cs to extract 1-D temporal features, as shown in Figure 5 (A), in two phases:

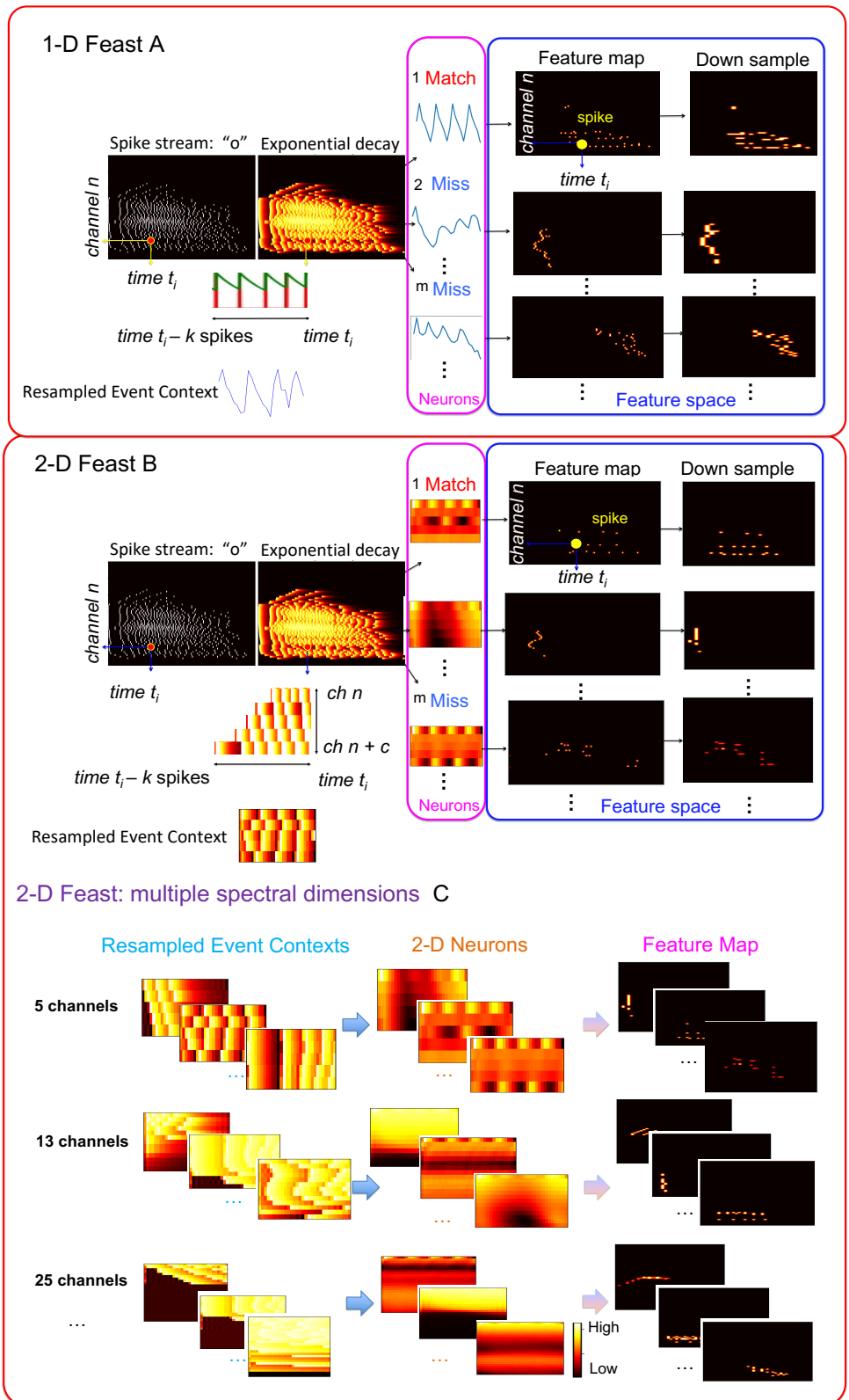

Figure 5 Event-based Spectrotemporal Feature Extraction: (A) 1-D FEAST; (B) 2-D FEAST; (C) 2-D FEAST on multiple spectral dimensions.

a) Learning:
In the learning phase, the number of the neurons, *m*, is pre-set, and the initial threshold and weights for each neuron are randomly generated. For each event, we choose *k* spikes in the past that are the closest to it and resample it to form its E_C. All the extracted E_Cs are presented in random order during training.

For an event at time $t_i$ and *channel n*, the dot product between its E_C and each neuron is calculated. The only neuron with the largest value which is also above its threshold is the winner. The threshold of the winner neuron is then increased by $\Delta I$, and the weights are updated according to (7). If there is no winner, all the neurons' thresholds are then decreased by a fixed amount of $\Delta E$. Multiple epochs of learning are performed empirically until it is converged.

b) Feature Extraction:
Once the system is no longer learning, the *m* neurons are then used to extract features from spike streams. Each neuron generates a feature map in its feature space: For an event at time $t_i$ in channel *n*, the dot product between its E_C and each neuron is calculated. The only neuron with the largest value is the winner. The winner neuron will emit a spike at time $t_i$ in channel *n* in its feature space to form a feature map.

The 1-D FEAST extracts channel-wised temporal features, in particular the slow changes in amplitude encoded in the spike streams. It is comparable to the computational spectrotemporal cortical model (Chi et al., 2005) that uses slow rate filters for the temporal analysis to extract syllabic rates in speech.

### 2.2.3 Spectro-temporal Feature Extraction – 2-D FEAST

Speech contains spectral modulations created by harmonics and formants, which are also evident in the cochleogram. Harmonics come from the vocal folds and are considered the source of the sound. Formants come from the vocal tract. Formants filter the harmonic sound source, and thus after harmonics go through the vocal tract, some become louder, and some become softer.

The features of the harmonics/formants are assosiated with the frequency channels and the next step of the Event-based Spectrotemporal Feature Extraction, 2-D FEAST, is to extract spectral and temporal combined features. As shown in Figure 5 (B), to extend the 1-D E_C, for each event at *time $t_i$* in *channel n*, we choose *c* channels in frequency. Within each selected channel, *k* spikes that are closest to the current event are selected and resampled to form a 2-D E_C. Similar to the 1-D FEAST, in the learning phase, for an event at time $t_i$ and *channel n*, the dot product between its E_C and each 2-D neuron is calculated. The neuron with the largest value which is also above its threshold is the winner. The threshold of the winner neuron is then increased by ∆*I*, and the weights are updated according to (7). If there is no winner, all the neurons' thresholds are then decreased by a fixed amount of ∆*E*. Multiple epochs of learning are performed until the weights have converged. In the feature extraction phase, for each event at time $t_i$ and channel *n*, the dot product between its E_C and each 2-D neuron is calculated. The only neuron with the largest value is the winner. The winner neuron will emit a spike at time $t_i$ in channel *n* in its feature space to form a feature map.

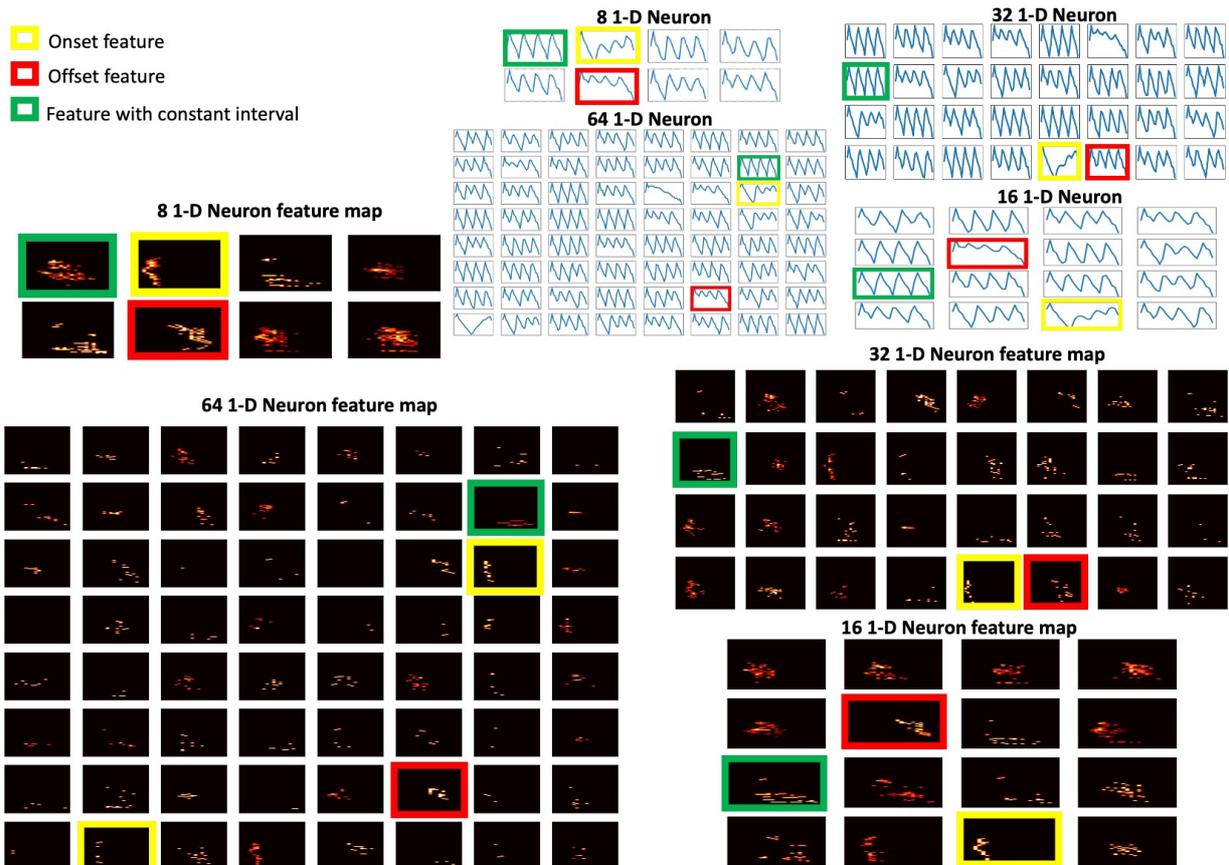

Figure 6 The 1-D neuron features with different configurations and the corresponding feature maps.

Furthermore, for each event, we generate multiple sets of 2-D E_Cs, as shown in Figure 5 (C). Each set includes a different number of channels so that it covers multiple scales in frequency. For example, as shown in Figure 5 (C), a 5-channel dimension only includes one harmonic, whereas a 13-channel can cover two harmonics, and so on. The choice of channel numbers is based on the Greenwood function used in the CAR-FAC model (Greenwood, 1990). We then apply the 2-D FEAST described previously on each set of the E_C in parallel.

The same learning and feature extraction phases described above are applied to each event. For each event at time $t_i$ and *channel n*, there is one winner neuron in each dimension. The winner neuron of each dimension will emit a spike at time $t_i$ in channel *n* in its feature space.

2-D FEAST is comparable to a spectrotemporal cortical model that uses different "seed functions" as scale filters for spectrotemporal analysis (Chi et al., 2005) to extract harmonics and formants.

## 3 RESULTS

### 3.1 CAR-FAC ON FPGA

The CAR-FAC FPGA implementation has been investigated and measured by (Xu et al., 2016) and (Xu, Thakur, et al., 2018). In this work, we use the proposed CAR-FAC cochlear system on FPGA to generate spike streams from the TIDIGITS database. Here $\tau_{LIF}$ in (2) is set as 10 ms, $V_{reset}$ is set as 0, and $threshold$ is only set as a medium value, 0.0004. The device utilisation of the binaural 2×64×9 CAR-FAC system is shown in Table1.

Table1 Device utilisation summary

|  | Used | Available | Utilisation |
|---|---|---|---|
| **ALM** | 13,958 | 29,080 | 48% |
| **Memory (bits)** | 1,278,771 | 4,567,040 | 28% |
| **DSPs** | 123 | 150 | 82% |

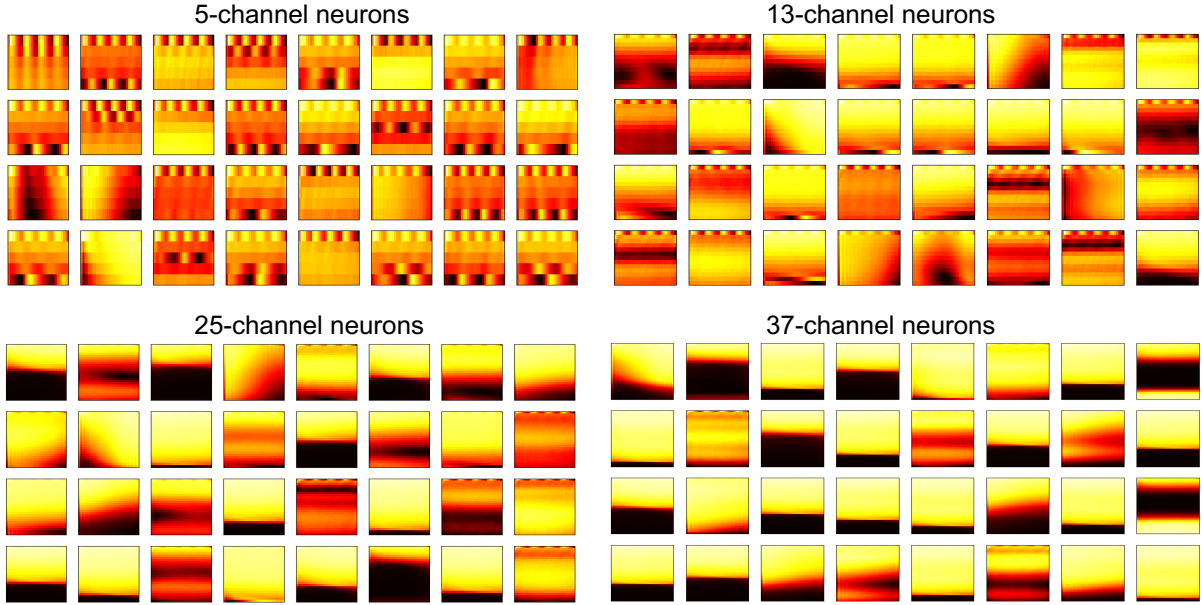

Figure 7 32-neuron features for all the different 2-D neuron with different dimensions.

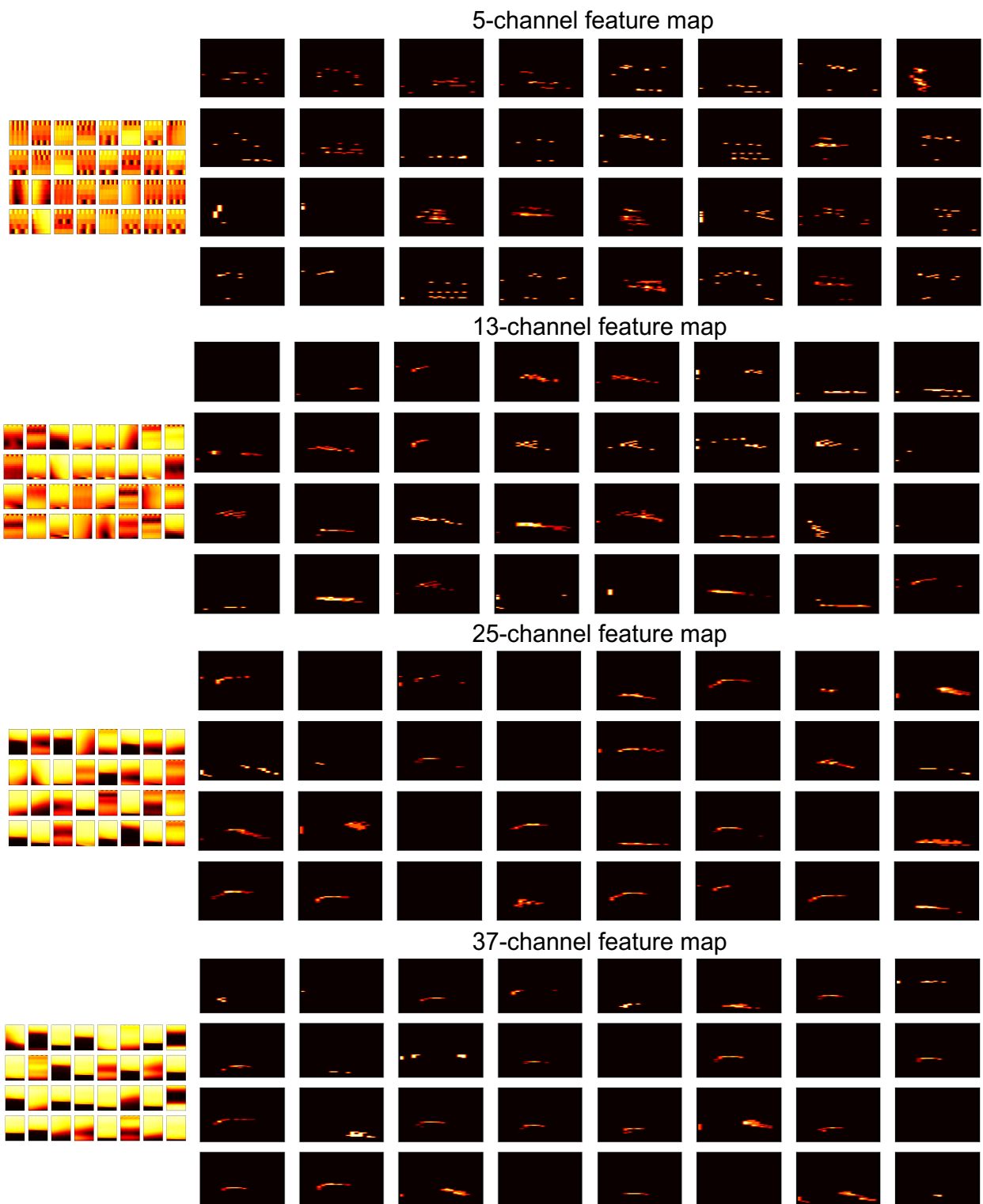

Figure 8 The 32 2-D neurons and down-sampled feature maps.

Table 2 Summary of investigated features on the TIDIGITS dataset

| Pre-processing | Method | Neuron number | Accuracy |
|---|---|---|---|
| CAR-FAC | Time-binned feature + SVM (baseline) | NA | **88.49%** |
| | 1-D FEAST + SVM | 8 | **92.96%** |
| | | 16 | **93.56%** |
| | | 32 | **93.92%** |
| | | 64 | **93.64%** |
| | 2-D FEAST + SVM | 16 (5ch) | **93.56%** |
| | | 16 (13ch) | **95.78%** |
| | | 16 (25ch) | **96.34%** |
| | | 16 (37ch) | **95.65%** |
| | | 16×4 (5+13+25+37ch) | **97.47%** |
| | | 32 (5ch) | **95.01%** |
| | | 32 (13ch) | **96.62%** |
| | | 32 (25ch) | **97.42%** |
| | | 32 (37ch) | **96.70%** |
| | | 32×4 (5+13+25+37ch) | **97.71%** |
| | | 64 (5ch) | **94.97%** |
| | | 64 (13ch) | **96.13%** |
| | | 64 (25ch) | **97.22%** |
| | | 64 (37ch) | **96.82%** |
| | | 64×4 (5+13+25+37ch) | **97.43%** |
| AMS1c (Anumula et al., 2018) | Constant time bins: Exponential features (GRU RNN with Backpropagation) | NA | 91.1% |
| MFCC | MFCC-SOM SLAYER with Backpropagation (Shrestha & Orchard, 2018) | 484-500-500-11 | 99.09% |

### 3.2 FEAST AND LINEAR CLASSIFIER

The 1-D and 2-D FEAST are tested, respectively, on an isolated spoken digit recognition task using the TIDIGIT dataset. Here we use the isolated spoken digits (zero to ten) from 225 speakers (female and male) as the training and testing data, of which 4950 samples are included in the total (2464 for training and 2486 for testing). The Support Vector Machine (SVM) with linear kernels and optimal regularisation is used as the back-end classifier to investigate the performance of the FEAST.

#### 3.2.1 1-D feature for temporal feature extraction

In the 1-D FEAST, we chose $k = 4$ spikes for an E_C. The E_C is then resampled into 32 samples. In this experiment, the algorithm had converged after ten epochs of training. The parameters were configured as $\Delta I = 0.001$, $\Delta E = 0.003$, and $\eta = 0.001$ in (7), which were derived empirically. The optimal number of neurons depend greatly on the nature of the data. In this experiment, 8, 16, 32, and 64 neurons are tested. The generated feature map for each neuron is down-sampled via fixed time binning (Anumula et al., 2018), as shown in Figure 6. By observing the features of the neurons, we can see the neuron with gradually decreasing intervals often represents an onset, whereas the neuron with gradually increasing intervals represents an offset of an utterance. The evenly distributed intervals represent an unchanged amplitude of the utterance. The generated 1-D features are then used as input for the SVM. Additionally, according to (Acharya et al. 2018) and (Anumula et al., 2018), the time-binned spikes show the highest accuracy compared to other statistical features in the isolated spoken digit recognition, so in this experiment, the time-binned spikes generated from the proposed cochlear system are investigated as a baseline. The classification results are shown in TABLE 2. For all the configurations, the 1-D FEAST shows better accuracy than the time-binned spikes, and the 32-neuron configuration shows the best accuracy, 93.92%.

#### 3.2.2 2-D feature for temporal feature extraction

In the 2-D FEAST, we chose $k = 4$ spikes and resample them to 32 samples temporally, and 5, 13, 25, and 37 channels for the 2-D E_Cs. In the training phase, we train each set of the 2-D E_Cs in parallel, using 16, 32 and 64 neurons, respectively. Figure 7 shows all the features of the 32 neurons and Figure 8 shows the corresponding feature maps. The small-sized neurons tend to show fine spectral features, whereas the large-sized neurons only show coarse intensity information in Figure 8.

We then use the 2-D features as input for the SVM. Firstly, we test each set of neurons separately. As shown in Table 2, the 32 neuron system shows the best accuracy, and we found for the same number of neurons, the 25-channel size tends to provide better accuracy. Next, we combine all the sizes together for each neuron configuration, and get an improved accuracy, 97.71%. As comparisons, the results of the same experiment by (Anumula et al., 2018) are also shown in TABLE 2, in which a Gated Recurrent Unit (GRU) Recurrent Neural Network (RNN) is used for a constant time binning of the exponential features. Currently, the highest accuracy of 99.09% on the same task is achieved by (Shrestha & Orchard, 2018) using a 484-500-500-11 neuron spiking neural network with backpropagation, whereas in our approach, we only use one-layer of 128 neurons (32 neurons × 4 sizes) and a simple linear classifier.

4 DISCUSSIONS

This paper presents a reconfigurable digital implementation of an event-based binaural cochlear system and an event-driven spectrotemporal receptive field feature extraction approach. The algorithm is tested on an isolated spoken digit recognition task. The features extracted from FEAST provide better multi-resolution representations of the event-based data than statistical approaches that have been classically used for decoding spike streams.

Like any other data modalities, noise in event-data poses challenges to effective processing and FEAST helps in learning noise-robust features. The CAR-FAC model has been shown to provide noise-robust features in audio to perform speaker identification (Islam, Xu, Monk, Afshar, & van Schaik, 2022). Audio features from the CAR-FAC cochlea model have also been used to perform noise-robust binaural sound localisation (Xu, Afshar, et al., 2018; Xu et al., 2019, 2021).

Since the FEAST is an unsupervised method, it cannot perform classification and requires a backend classifier. In follow-up work, we will use a generalised model of the FEAST method that performs feature extraction and classification in a single architecture (Bethi et al., 2022).

5 REFERENCES

Afshar, S., George, L., Thakur, C. S.,vTapson, J., van Schaik, A., de Chazal, P., & Hamilton, T. J. (2015). Turn Down That Noise: Synaptic Encoding of Afferent SNR in a Single Spiking Neuron. *IEEE Transactions on Biomedical Circuits and Systems*, *9*(2), 188–196. https://doi.org/10.1109/TBCAS.2015.2416391

Afshar, S., Hamilton, T. J., Davis, L., van Schaik, A., & Delic, D. (2020). Event-Based Processing of Single Photon Avalanche Diode Sensors. *IEEE Sensors Journal*, *20*(14), 7677–7691. https://doi.org/10.1109/JSEN.2020.2979761

Afshar, S., Nicholson, A. P., van Schaik, A., & Cohen, G. (2020). Event-Based Object Detection and Tracking for Space Situational Awareness. *IEEE Sensors Journal*, *20*(24), 15117–15132. https://doi.org/10.1109/JSEN.2020.3009687

Afshar, S., Ralph, N., Xu, Y., Tapson, J., van Schaik, A., & Cohen, G. (2020). Event-based feature extraction using adaptive selection thresholds. *Sensors (Switzerland)*, *20*(6). https://doi.org/10.3390/s20061600

Anumula, J., Neil, D., Delbruck, T., & Liu, S. C. (2018). Feature representations for neuromorphic audio spike streams. *Frontiers in Neuroscience*, *12*(FEB). https://doi.org/10.3389/fnins.2018.00023

Baldwin, R., Liu, R., Almatrafi, M. M., Asari, V. K., & Hirakawa, K. (2022). Time-Ordered Recent Event (TORE) Volumes for Event Cameras. *IEEE Transactions on Pattern Analysis and Machine Intelligence*, *14*(8), 1–14. https://doi.org/10.1109/TPAMI.2022.3172212

Bethi, Y., Xu, Y., Cohen, G., van Schaik, A., & Afshar, S. (2022). An Optimized Deep Spiking Neural Network Architecture Without Gradients. *IEEE Access*, *10*(September), 97912–97929. https://doi.org/10.1109/ACCESS.2022.3200699

Chakrabartty, S., & Liu, S. C. (2010). Exploiting spike-based dynamics in a silicon cochlea for speaker identification. *ISCAS 2010 - 2010 IEEE International Symposium on Circuits and Systems: Nano-Bio Circuit Fabrics and Systems*, 513–516. https://doi.org/10.1109/ISCAS.2010.5537578

Chi, T., Ru, P., & Shamma, S. A. (2005). Multiresolution spectrotemporal analysis of complex sounds. *The Journal of the Acoustical Society of America*. https://doi.org/10.1121/1.1945807

Cohen, G., Afshar, S., Morreale, B., Bessell, T., Wabnitz, A., Rutten, M., & van Schaik, A. (2019). Event-based Sensing for Space Situational Awareness. *Journal of the Astronautical Sciences*, *66*(2), 125–141. https://doi.org/10.1007/s40295-018-00140-5

Gerstner, W., & Kistler, W. M. (2002). Spiking Neuron Models: Single Neurons, Populations, Plasticity. *Book*, 494. https://doi.org/10.2277/0511075065

Greenwood, D. D. (1990). A cochlear frequency-position function for several species - 29 years later. *The Journal of the Acoustical Society of America*, Vol. 87, pp. 2592–2605. https://doi.org/10.1121/1.399052

Haessig, G., Milde, M. B., Aceituno, P. V., Oubari, O., Knight, J. C., van Schaik, A., … Indiveri, G. (2020). Event-Based Computation for Touch Localization Based on Precise Spike Timing. *Frontiers in Neuroscience*, *14*(May), 1–19. https://doi.org/10.3389/fnins.2020.00420

Islam, M. A., Xu, Y., Monk, T., Afshar, S., & van Schaik, A. (2022). Noise-robust text-dependent speaker identification using cochlear models. *The Journal of the Acoustical Society of America*, *151*(1), 500–516. https://doi.org/10.1121/10.0009314

Lagorce, X., Ieng, S. H., Clady, X., Pfeiffer, M., & Benosman, R. B. (2015). Spatiotemporal features for asynchronous event-based data. *Frontiers in Neuroscience*. https://doi.org/10.3389/fnins.2015.00046

Li, C. H., Delbruck, T., & Liu, S. C. (2012). Real-time speaker identification using the AEREAR2 event-based silicon cochlea. *ISCAS 2012 - 2012 IEEE International Symposium on Circuits and Systems*, 1159–1162. https://doi.org/10.1109/ISCAS.2012.6271438

Liu, S. C., van Schaik, A., Minch, B. A., & Delbruck, T. (2010). Event-based 64-channel binaural silicon cochlea with Q enhancement mechanisms. *ISCAS 2010 - 2010 IEEE International Symposium on Circuits and Systems: Nano-Bio Circuit Fabrics and Systems*. https://doi.org/10.1109/ISCAS.2010.5537164

Liu, S. C., van Schaik, A., Minch, B. A., & Delbruck, T. (2014). Asynchronous binaural spatial audition sensor with 2×64×4 Channel output. *IEEE Transactions on Biomedical Circuits and Systems*. https://doi.org/10.1109/TBCAS.2013.2281834

Lyon, R. F. (2017). *Human and Machine Hearing -Extracting Meaning from Sound*. Cambridge University Press.

Maqueda, A. I., Loquercio, A., Gallego, G., Garcia, N., & Scaramuzza, D. (2018). Event-Based Vision Meets Deep Learning on Steering Prediction for Self-Driving Cars. *Proceedings of the IEEE Computer Society Conference on Computer Vision and Pattern Recognition*, (Dl), 5419–5427. https://doi.org/10.1109/CVPR.2018.00568

Neil, D., & Liu, S. C. (2016). Effective sensor fusion with event-based sensors and deep network architectures. *Proceedings - IEEE International Symposium on Circuits and Systems*, *2016-July*, 2282–2285. https://doi.org/10.1109/ISCAS.2016.7539039

Ralph, N., Joubert, D., Jolley, A., Afshar, S., Tothill, N., van Schaik, A., & Cohen, G. (2022). Real-Time Event-Based Unsupervised Feature Consolidation and Tracking for Space Situational Awareness. *Frontiers in Neuroscience*, *16*(May), 1–26. https://doi.org/10.3389/fnins.2022.821157

Rasetto, M., Dominguez-Morales, J. P., Jimenez-Fernandez, A., & Benosman, R. (2021). *Event Based Time-Vectors for auditory features extraction: a neuromorphic approach for low power audio recognition*. Retrieved from http://arxiv.org/abs/2112.07011

Shrestha, S. B., & Orchard, G. (2018). Slayer: Spike layer error reassignment in time. *Advances in Neural Information Processing Systems*, *2018-Decem*(NeurIPS), 1412–1421.

Singh, R. K., Xu, Y., Wang, R., Hamilton, T. J., van Schaik, A., & Denham, S. L. (2018). CAR-Lite: A Multi-Rate Cochlear Model on FPGA for Spike-based Sound Encoding. *IEEE Transactions on Circuits and Systems I: Regular Papers*, 1–13. https://doi.org/10.1109/ISCAS.2018.8351394

Tapson, J., Cohen, G., & van Schaik, A. (2015). ELM solutions for event-based systems. *Neurocomputing*, *149*(Part A), 435–442. https://doi.org/10.1016/j.neucom.2014.01.074

Uysal, I., Sathyendra, H., & Harris, J. G. (2006). A biologically plausible system approach for noise robust vowel recognition. *Midwest Symposium on Circuits and Systems*, *1*, 245–249. https://doi.org/10.1109/MWSCAS.2006.382043

Xu, Y. (2019). *A Digital Neuromorphic Auditory Pathway*. WESTERN SYDNEY UNIVERSITY.

Xu, Y., Afshar, S., Singh, R. K., Hamilton, T. J., Wang, R., & van Schaik, A. (2018). A Machine Hearing System for Binaural Sound Localization based on Instantaneous Correlation. *2018 IEEE International Symposium on Circuits and Systems (ISCAS)*. https://doi.org/10.1109/ISCAS.2018.8351367

Xu, Y., Afshar, S., Singh, R. K., Wang, R., van Schaik, A., & Hamilton, T. J. (2019). A binaural sound localization system using deep convolutional neural networks. *Proceedings - IEEE International Symposium on Circuits and Systems*. https://doi.org/10.1109/ISCAS.2019.8702345

Xu, Y., Afshar, S., Wang, R., Cohen, G., Thakur, C. S., Hamilton, T. J., & Van Schaik, A. (2021). A biologically inspired sound localisation system using a silicon cochlea pair. *Applied Sciences (Switzerland)*, *11*(4), 1–21. https://doi.org/10.3390/app11041519

Xu, Y., Thakur, C. S., Singh, R. K., Hamilton, T. J., Wang, R. M., & van Schaik, A. (2018). A FPGA implementation of the CAR-FAC cochlear model. *Frontiers in Neuroscience*, *12*(APR), 1–14. https://doi.org/10.3389/fnins.2018.00198

Xu, Y., Thakur, C. S., Singh, R. K., Wang, R., & van Schaik, A. (2016). Electronic Cochlea: CAR-FAC Model on FPGA. *IEEE Biomedical Circuits and Systems Conference*, 1–4.

Yang, M., Chien, C.-H., Delbruck, T., & Liu, S.-C. (2016). A 0.5 V 55 μW 64 × 2 Channel Binaural Silicon Cochlea for Event-Driven Stereo-Audio Sensing. *IEEE Journal of Solid-State Circuits*, *51*(11), 2554–2569. https://doi.org/10.1109/JSSC.2016.2604285

Zappa, F., Villa, F., Lussana, R., Delic, D., Mau, M. C. J., Redouté, J. M., … Afshar, S. (2020). Microelectronic 3D imaging and neuromorphic recognition for autonomous UAVs. In C. Palestini (Ed.), *Advanced Technologies for Security Applications* (pp. 185–194). https://doi.org/10.1007/978-94-024-2021-0_17